\documentclass[11pt]{article}
\usepackage{amsmath,amssymb,mathdots}
\usepackage{latexsym}
\usepackage{datetime}
\usepackage[all]{xy}

\def\e#1\e{\begin{equation}#1\end{equation}}
\def\ea#1\ea{\begin{align}#1\end{align}}

\def\narrow{\par\addtolength{\leftskip}{20pt}\addtolength{\rightskip}{5pt}}
\def\oldmargins{\par\setlength{\leftskip}{0pt}\setlength{\rightskip}{0pt}}
\newcounter{quest}[section]
\newcounter{qa}[quest]
\newcounter{qi}[quest]

\def\inext{\par\ifnum\value{qi}=0 \narrow\fi \medskip\addtocounter{qi}{1}
\noindent\hbox to 0pt{\hss\bf(\roman{qi})\hskip .3em}}
\def\anext{\par\ifnum\value{qa}=0 \narrow\fi \medskip\addtocounter{qa}{1}
\noindent\hbox to 0pt{\hss\bf(\alph{qa})\hskip .3em}}

\numberwithin{equation}{section}

\newcommand{\be}{\begin{equation}}
\newcommand{\ee}{\end{equation}}
\newcommand{\bes}{\begin{equation*}}
\newcommand{\ees}{\end{equation*}}
\newcommand{\bea}{\begin{eqnarray}}
\newcommand{\eea}{\end{eqnarray}}

\def\w{\wedge}

\def\pa{\partial}
\def\bpa{{\bar \partial}}

\def\ds{d^*}
\def\dl{d^\Lambda}

\def\dc{d^c}

\def\Om{\Omega}

\def\CB{\mathcal{P}}
\def\CH{\mathcal{H}}

\def\CU{\mathcal{U}}

\def\A{\mathcal{A}}

\def\La{\Lambda}

\def\d{\delta}
\def\la{\lambda}
\def\om{\omega}

\def\im{{\rm im~}}

\def\Re{{\rm Re}\,}
\def\Im{{\rm Im}\,}
 
\def\ea{e_1}

\def\dpp{{\partial_+}}
\def\dpm{{\partial_-}}
\def\dppm{{\partial_{\pm}}}

\def\dJ{{d^{\mathcal J}}}

\def\cj{\mathcal{J}}
\def\cja{{\mathcal{J}_1}}
\def\cjb{{\mathcal{J}_2}}

\begin{document}



\title{\bf{Generalized Cohomologies and Supersymmetry
}}

\author{Li-Sheng Tseng$^1$ and Shing-Tung Yau$^2$\\
~~\\
$^1$Department of Mathematics, University of California, Irvine, CA 92697 USA\\
{\tt lstseng@math.uci.edu}\\
~~\\
$^2$Department of Mathematics, Harvard University, Cambridge, MA 02138 USA\\
{\tt yau@math.harvard.edu}\\
~~\\
}

\date{\usdate{\today}}

\maketitle

\begin{abstract}\
We show that the complex cohomologies of Bott, Chern, and Aeppli and the symplectic cohomologies of Tseng and Yau arise in the context of type II string theory.  Specifically, they can be used to count a subset of scalar moduli fields in Minkowski compactification with RR fluxes in the presence of either $O5/D5$ or $O6/D6$ brane sources, respectively.  Further, we introduce a new set of cohomologies within the generalized complex geometry framework which interpolate between these known complex and symplectic cohomologies.  The generalized complex cohomologies play the analogous role for counting massless fields for a general supersymmetric Minkowski type II compactification with Ramond-Ramond flux.

\end{abstract}


\thispagestyle{empty}

\newpage

\setcounter{equation}{0}
\setcounter{footnote}{0}
\setcounter{page}{1}

\section{Introduction}

A basic and important question for supersymmetric flux compactification in string theory is to determine the number of scalar massless fields, or equivalently moduli fields, for a generic background solution.  Except for special flux solutions where the underlying manifold is Calabi-Yau or dual to one that is Calabi-Yau\footnote{In this paper, Calabi-Yau refers to the existence of a K\"ahler Calabi-Yau metric.}, there is at present no known systematic way to count scalar moduli fields when the solution manifold is non-K\"ahler.   Geometrically, the procedure to understand the scalar moduli starts with the linearized variation of the supersymmetry equations of supergravity.  In the Calabi-Yau case, the physically distinct solutions of the linearized equations can be nicely parametrized by the Dolbeault harmonic forms and counted by the associated Hodge numbers (see e.g. \cite{CO}).  In the general non-K\"ahler case, the linearized supersymmetric equations were written down in \cite{BTY} for the heterotic case and in \cite{Tomasiello} for the type II case.  However, the system of linearized equations for a general flux background is sufficiently complicated that it is challenging to find a straightforward interpretation of the full solution space.

In this paper, we take a modest step and ask whether there exists a subset of the linearized solution space that is more tractable and easier to characterize.  We focus on Minkowski $M^{3,1}\times X^6$ supersymmetric solutions of type II strings and its linearized system of equations.   Imposing some simplifying conditions on the linearized system, we show that indeed a subspace of the linearized solution space can be parametrized by harmonic elements of certain cohomologies of differential forms.  In the type IIB case with $O5/D5$ brane sources, the internal compact manifold $X^6$ is complex and we find that the complex cohomologies introduced by Bott and Chern \cite{BC}, and  Aeppli \cite{Aeppli}, can be used to count a subset of the massless deformations.  These cohomologies are isomorphic to the Dolbeault cohomology on a K\"ahler manifold.  However, for a general complex non-K\"ahler manifold, especially when the $\pa\bpa$-lemma does not hold, Bott-Chern and Aeppli cohomologies can have different dimensions as compared to the Dolbeault cohomology and encode different complex invariants.  Separately, in the type IIA case with $O6/D6$ brane sources, which requires that $X^6$ be a symplectic manifold, the cohomologies of interest turn out to be two that were recently introduced by Tseng and Yau \cite{TY1, TY2}.  
And in the more general supersymmetric type II string background with orientifold and D-brane sources, the supersymmetric equations stipulate that $X^6$ is generalized complex \cite{GMPT,GMPT1}.   Hence, ``generalizing" the complex cohomologies of Bott-Chern and Aeppli  and the symplectic cohomologies of Tseng-Yau, we are led to two new generalized complex cohomologies that interpolate between them.
The generalized complex cohomologies can be used to count a subset of massless fields in a general type II Minkowski background with Ramond-Ramond fluxes.

\section{Supersymmetry equations and cohomology}

This work on cohomology of type II strings can be motivated in part by noting certain similarities between the Maxwell equations and the  type II Minkowski $N=1$ supergravity equations, as written in the generalized complex form by Grana-Minasian-Petrini-Tomasiello \cite{GMPT1} and Tomasiello \cite{Tomasiello}.  As the solution space of the Maxwell equations is intrinsically linked with the de Rham cohomology, one can ask whether any cohomology is suggested by the supersymmetric type II equations.  Of course, the type II equations are gravitational in nature and highly non-linear and a priori one should not expect any simple cohomology to come out of them.  But it turns out that if we are willing to impose certain constraints on the solution space of the type II equations, then the type II differential system can be studied in an analogous manner with that of the Maxwell equations.

Let us begin by first recalling one connection of the Maxwell equations with the de Rham cohomology.  The Maxwell equations on some four-manifold $(X^4, g)$ has the simple form: 
\begin{align}
d \,F&= 0 \label{ecc} \\
d * F &= \rho_e \label{ecd} 
\end{align}
where $F$ is the curvature two-form of a $U(1)$ bundle, $\rho_e$ is the Poincar\'e dual three-current of a configuration of electric source particles.  (An electric particle maps out a one-dimensional worldline in $X^4$ so its Poincar\'e dual is a three-current.)  In this setting, if we want to consider the moduli space of solutions for $F$ in a fixed charge configuration (i.e. with $\rho_e$ fixed), then the variation of $F \to F + \d F$ implies that $\d F$
satisfies
\be\label{Mcond}
d\,\d F =0 ~, \qquad\qquad d^*\, \d F = 0~, 
\ee
which are the harmonic conditions of the de Rham class $H^2(X^4)\,$.
Hence,  the solution space of $F$ with $\d \rho_e$ fixed (i.e. $\d\rho_e=0$) is parametrized by de Rham harmonic two-forms, $\CH^2(X^4)\,$.

The type II supergravity equations of our focus are those that arise from imposing ${N=1}$ supersymmetry on $M^{3,1} \times X^6\,$ - the product of Minkowski spacetime and a compact six-dimensional manifold - with the conformally warped metric 
$$ ds^2 = e^{2f}\,ds^2_{M^{3,1}}
 + ds^2_{X^6}~,$$
with $e^{2f}$ being the conformal factor.   The supersymmetric equations can be written simply in the generalized complex geometry framework \cite{GMPT1,Tomasiello}.  Below, instead of jumping directly into the generalized complex equations, we shall build up our intuition by first examining the special case in type IIB theory where solutions on $X^6$ have the more familiar $SU(3)$ structure and are complex.  Then we shall turn to the symplectic solutions in type IIA theory also with an $SU(3)$ structure.   With the special cases worked out, we will finally turn to the most general solutions with RR-flux which have an $SU(3)\times SU(3)$ structure and are generalized complex.

\subsection{Complex cohomology in type IIB supergravity solutions with $O5$-brane}

Type IIB superysmmetric solutions with $O5/D5$-brane sources are required to be complex.  Such solutions have an $SU(3)$ structure which is encoded in the hermitian $(1,1)$-form $\om$ and a non-where vanishing decomposable $(3,0)$-form $\Om$ on $X^6$.  The $SU(3)$ data $(\Om, \om)$ satisfy the algebraic conditions:
\begin{align}
  \om \w \Om &=0 \label{eca}~,\\
 i\,\Omega \w {\bar \Omega} &= 8 \,e^{2f}\,\frac{\om^3}{3!}~. \label{ecaa}
\end{align}
Moreover, they satisfy the following differential conditions:
\begin{align}
d\Omega&=0~, \label{ecab}\\
d(\om^2/2)&=0~,\label{ecac}\\
d\dc(e^{-2f} \om)& = \rho_B~, \label{ecad}
\end{align}
where the differential operator $\dc= i \, (\bpa-\pa)$ (hence, $d\dc = 2i\,\pa\bpa$) and $\rho_B$ is the Poincar\'e dual four-current of holomorphic submanifolds (of complex codimension two) which the orientifold five-branes and/or D5-branes wrap around.  The above system of equations modulo a conformal rescaling is just a special case of the general system of equations written in \cite{Tomasiello}.  In this form, the system contains only geometrical quantities $(\Om, \om, \rho_B)$ plus the conformal factor $e^{-2f}$.  Though non-trivial background solutions of this differential system do have a non-zero three-form flux $F_3$, this flux form has been implicitly solved by imposing supersymmetry:
$$F_3 = d^c ( e^{-2f} \omega)~.$$
Indeed, the last equation \eqref{ecad} is simply the magnetic source equation $dF_3 = \rho_B$.

Having presented the equations in the above form, we can quickly see a resemblance to the Maxwell equations, in particular for \eqref{ecac}-\eqref {ecad}.  Using the relation $\om = * (\om^2/2)\,$, where the Hodge star is defined with respect to the compatible metric defined by $\om$, equations \eqref{ecac}-\eqref{ecad} can be re-written as
\begin{align}
d(\om^2/2)&=0 \label{ecae}\\
d\dc [e^{-2f}* (\om^2/2)]& = \rho_B \label{ecaf}
\end{align}
which (up to the conformal factor $e^{-2f}$) has the form of Maxwell's equations \eqref{ecc}-\eqref{ecd}, identifying $F_2 \sim \om^2/2$, $d* \sim d\dc*$, and $\rho_e \sim \rho_B$.

Let us now study the local solution space by considering a linearized deformation of this type IIB system of equations on a fixed complex manifold (specifically keeping $\Omega$ fixed).  We would like here to mirror the analysis of the Maxwell case, so we will keep the source current fixed (i.e. $\delta \rho_B= 0$).  Additionally, we will impose the constraint that the conformal factor does not vary (i.e. $\delta f =0$).  This is a useful simplifying condition as $e^{-2f}$ not only appears in \eqref{ecaf} but is also dependent on $\om$ through \eqref{ecaa}.  Performing a linearized variation $\om \to \om + \delta \om\,$ in \eqref{ecaa}, we see that $\delta f=0$ (and $\delta \Omega =0$) imposes  
$$\om^2 \w \delta \om = 0$$
which is the condition that $\delta \om$ is a primitive form.\footnote{In dimension $d=2n$, a differential $k$-form $B_k$ with $k\leq n$ is called primitive if $\om^{n-k+1}\w B_k = 0\,$.}  Coupled with the variation of  \eqref{eca} which gives the condition 
$$\Om \w \delta \om = 0~,$$
we can conclude that the linearized variation $\delta \om$ must be a primitive (1,1)-form if  we impose $\delta\Omega= \delta \rho_B = \delta f =0$ on the above type IIB system.

With the specified constraints, the linearized deformation of the type II system effectively reduces down to just deforming the hermitian form $\om$ by a primitive (1,1)-form in equations  \eqref{ecac}-\eqref{ecad}.  Now, the Hodge star of a primitive (1,1)-form $\delta \om$ has a simple expression and can be written as (see e.g. \cite{Huy})
\be\label{Bdual}
\delta \om = - *(\om \w \delta \om) = - * \delta(\om^2/2)~.
\ee
Thus, the linearized deformed equations \eqref{ecac}-\eqref{ecad}
gives us the conditions for $\delta(\om^2/2)$
\be \label{bcond} d \, \delta(\om^2/2) =0~,\qquad d\dc\, e^{-2f}* \delta(\om^2/2) =0~.
\ee

The conditions in \eqref{bcond} are similar to that of the Maxwell case in \eqref{Mcond}.  In fact, if we multiply the second equation of \eqref{bcond} on the left by $e^{2f}*$, then it becomes $(d\dc)*'\delta(\om^2/2)=0\,$, where the adjoint operator $(d\dc)*'$ is defined with respect to a conformally weighted inner product. This close similarity with the Maxwell case then begs the question whether there is a cohomology whose harmonic forms are $d$- and $(d\dc)^*$-closed.  Indeed, these are precisely the conditions for the harmonic forms of the so-called Bott-Chern cohomology \cite{BC, Aeppli},
$$H_{BC}^{p,q}(X) = \frac{\{A^{p,q}\in \A^{p,q}| ~d\, A^{p,q} =0\}}{d\dc\,\A^{p-1,q-1}}$$
where $\A^{p,q}$ is the space of $(p,q)$-forms.
Thus, having imposed the conditions $\delta\Omega= \delta \rho_B = \delta f =0$, we find that the linearized deformation $\delta(\om^2/2)= \om \w \delta \om$, is parametrized by
$$\delta(\om^2/2) \in \CH^{2,2}_{BC}(X)\cap (\om \w \CB^2)$$
where $\CB^2$ denotes the space of primitive 2-forms.

Alternatively, we can use \eqref{Bdual} to re-write the conditions of \eqref{bcond} directly in terms of $\delta \om$.  This gives
\be \label{bcondd} d * \delta\om =0~,\qquad d\dc\, e^{-2f} \delta\om =0~.
\ee
After rescaling $\delta \om\to e^{2f}\delta\om$, \eqref{bcondd} becomes a condition for a two-form that is $d\dc$- and $\ds$-closed.  Such exactly match the harmonic conditions of the Aeppli cohomology \cite{Aeppli}
$$H_{A}^{p,q}(X) = \frac{\{A^{p,q}\in \A^{p,q}|~ d\dc\, A^{p,q} =0\}}{\pa\,\A^{p-1,q} + \bpa\,\A^{p,q-1}}~.$$
Hence, an alternative way of parametrizing the deformation is 
$$\delta\om \in \CH^{1,1}_A(X) \cap \CB^2~.$$
And not surprisingly, it is possible to show that the harmonic forms of Bott-Chern and Aeppli cohomology are dual to each other, explicitly by the operation of the Hodge star operator (see e.g. \cite{Sch}).

\

\subsection{Symplectic cohomology in type IIA supergravity solutions with $O6$-brane}

Let us turn now to consider type IIA solutions with only $O6/D6$-branes sources.  The geometry on $X^6$ is symplectic with an $SU(3)$ structure.  Again, the $SU(3)$ geometrical data $(\om, \Om)$ satisfy the algebraic conditions:  
\begin{align}
  \om \w \Om &=0 ~,\label{esa}\\
8 \,\frac{\om^3}{3!} &=  i\,e^{2f} \,\Omega \w {\bar \Omega} ~. \label{esaa}
\end{align}
Note here that the conformal factor $e^{2f}$ in \eqref{esaa} is defined to be the inverse of that of the IIB complex system \eqref{ecaa}.  As for the symplectic differential equations, they take the form
\begin{align}
d\, \om & = 0 ~,\label{esab}\\
d\, {\rm Re}\, \Omega&=0~,\label{esac}\\
d\dl \,(e^{-2f} {\rm Im}\, \Omega)&=\rho_A ~,\label{esad}
\end{align}
where $\dl=d\La - \La d$ is the symplectic adjoint operator \footnote{The operation $\La$ is defined as the interior product with $\om^{-1}\,$.  Specifically, acting on a differential form $A$, $\La\, A = \frac{1}{2}(\om^{-1})^{ij}\,i_{\pa_{x^i}} i_{\pa_{x^j}} A\,$; hence, $\La$ is an operation that lowers the degree of differential forms by two.}, $\rho_A$ is the source term that is sourced by $O6$- and D6-branes wrapping special Lagrangian subspaces.   The RR-flux involved is the two-form $F_2$ which by supersymmetry is solved to be 
$$ F_2 = \dl \left(e^{-2f} {\rm Im}\, \Omega\right)~.$$
Again, the above equations up to a rescaling are just a special case of the general equations in \cite{Tomasiello}.  Further, if we note the relation,  ${\rm Im}\,\Omega = * {\rm Re}\, \Omega\,$, then \eqref{esac}-\eqref{esad} can be re-expressed as
\begin{align*}
d\, {\rm Re}\, \Omega&=0\\
d\dl \,e^{-2f} *({\rm Re}\, \Omega)&=\rho_A
\end{align*}
which motivate the comparison with the Maxwell equations \eqref{ecc}-\eqref{ecd}.

We perform now  a linearized variational analysis parallel to the IIB complex case in the previous subsection.  Treating the above equations \eqref{esa}-\eqref{esad} as a symplectic system, we shall consider the linearized deformation of the almost complex structure $\Om \to \Om + \delta \Om$ while imposing the following analogous conditions: (i)  the symplectic structure fixed, $\delta \om =0\,$; (ii) the source current fixed, $\delta \rho_A =0\,$; (iii) the conformal factor fixed, $\delta f =0\,$.

The variation of the first algebraic condition \eqref{esa} with $\delta \om = 0$ gives
$$\om \w \delta \Om =0~,$$
which implies that $\delta \Om$ is a primitive form.  The linearized variation of the second condition \eqref{esaa} gives
$$ \delta \Om \w {\bar \Omega} =0~,$$
thus further constraining $\delta \Om$ to be a primitive (2,1)-form.\footnote{In infinitesimally deforming the almost complex structure represented by $\Om\,$, $\delta \Om$ only has at most (3,0) and (2,1) components.}  Now for $\delta \Om$ that is a primitive (2,1)-form, we have
\be\label{Sdual}
\Im \delta\Om = \frac{1}{2i}\left(\delta \Om - \delta {\bar \Om}\right) = -\frac{1}{2} * \left(\delta \Om + \delta {\bar \Om}\right) = -* \Re \delta\Om~.
\ee
Hence, the linearized deformation of \eqref{esac}-\eqref{esad} with the imposed constraints give the conditions
\be\label{scond}
d\,\Re \d\Omega = 0~, \qquad d\dl e^{-2f} * \Re \d\Omega = 0~.
\ee
Multiplying the second equations by $e^{2f} *$, \eqref{scond} become the requirement that $\Re \d\Om$ is both $d$- and $(d\dl)*'$-closed, which are the harmonic conditions (of a conformally weighted inner product) of a primitive symplectic cohomology introduced by Tseng and Yau 
\cite{TY1,TY2}
$$PH^k_{d+\dl}(X)=\frac{\{B^k\in \CB^k| ~d\, B^k =0\}}{d\dl \CB^k}~.$$
So infinitesimally, we have that imposing $\d \om = \d \rho_A = \d f =0$, 
$$\Re \delta \Omega \in P\CH^3_{d+\dl} \cap \Re \A^{2,1}~.$$
with respect to the conformally weighted metric.

Alternatively, we can translate the result for $\Re \d \Omega$ into that for $\Im \d \Omega$.  Using \eqref{Sdual}, equation \eqref{scond} can be equivalently expressed as
\be\label{scondd}
d\dl e^{-2f} \Im \d \Om = 0 ~, \qquad d * \Im \d \Om =0~.
\ee
Up to a rescaling, this is just the harmonic condition for the dual primitive symplectic cohomology \cite{TY1,TY2}\footnote{$(\dpp,\dpm)$ are linear differential operators defined on symplectic spaces \cite{TY2}.  Acting on primitive forms, $\dppm: \CB^{k} \to \CB^{k\pm 1}$ and are defined simply as the projection of the exterior derivative operator onto the two primitive components.  Specifically, the action of $d$ on a primitive $k$-form $B_k\in\CB^k$ has only two terms under standard Lefschetz decomposition: $dB_k = B^0_{k+1} + \om \w B^1_{k-1}\,$, where $B^0$ and $B^1$ are also primitive forms.  Hence,   $\dpp B^{k} = B^0_{k+1}$ and $\dpm B = B^1_{k-1}\,$.} 
$$PH^k_{d\dl}(X) = \frac{\{B^k\in \CB^k|~ d\dl\, B^k = 0 \}}{\dpp \CB^{k-1}+ \dpm \CB^{k+1}}~.$$
Hence, we have
$$\Im \delta \Omega \in P\CH^3_{d\dl} \cap \Im \A^{2,1}~.$$

Let us add that since the O6/D6-branes wrap special Lagrangians subspaces which are defined by both $\Om$ and $\om$, the imposition of $\delta\rho_A=0$ in general might give an additional obstruction for the $\d \Omega$ variation.  An analogous obstruction does not arise in the complex case since a holomorphic submanifold is defined with respect to the complex structure only.

\subsection{Generalized cohomology for general type II supersymmetric supergravity solutions} 

Let us now turn to the general case of Minkowski compactification with RR flux in type II supergravity.  The background geometry on $X^6$ was found in \cite{GMPT} to have the generalized complex structure introduced by Hitchin \cite{Hitchin,Gua1, Caval}\footnote{Since there are now a number of expositions on generalized complex geometry (including some oriented for physicists, e.g. \cite{Zabzine, GMPT2, K2}), we refer the reader to the literature for background and standard notations.  Our conventions in this subsection mostly follow \cite{Tomasiello}. The difference in some signs and scale factors with other conventions in the literature does not factor in the identification of the cohomologies.}.  With an $SU(3)\times SU(3)$ structure on $TX \oplus TX^*\,$, $X^6$ has a pair of compatible almost generalized complex structures $(\cj_1,\cj_2)$ of which $\cja$ is integrable while $\cj_2$'s integrability fails when RR fluxes are present.  The supersymmetry equations can be expressed concisely in terms of the associated pure spinors $(\Phi_1, \Phi_2)$ \cite{GMPT1, Tomasiello}.  The compatible pure spinors are related by
\be\label{eqq}
\| \Phi_1\|^2 = e^{2f} \| \Phi_2 \|^2~.
\ee
and satisfy the following differential conditions on $X^6$:
\begin{align}
d\Phi_1 &= 0~, \label{eqa}\\
d \,{\rm Re}\, \Phi_2 &= 0 ~,\label{eqb}\\
dd^{\cja}\,(e^{-2f}\, {\rm Im}\, \Phi_2) & = \rho ~.\label{eqc}
\end{align} 
Here, the norm $\| \Phi \|^2$ is defined as the top form of the Mukai pairing
\be\label{mukai}
 \left(\Phi \w \la(\bar{\Phi})\right)_{top} = -i \| \Phi \|^2 ~vol~,
\ee
where $\la$ is an involutive operator whose action on a $k$-form is defined to be
$$\la(A_k) = (-1)^{k(k-1)/2} A_k~.$$
For a generalized complex structure $\cj$, $d^\cj$ is the operator defined by 
$$d^\cj = \cj^{-1} \, d\, \cj~. $$
In the complex case, $d^\cj= d^c$, while $d^\cj= \dl$ in the symplectic case.  
(We give a comparision of the generalized objects and their expression in the IIB complex system and the IIA symplectic system in Table \ref{firstops}.)  $\rho$ is the Poincar\`e dual of "generalized calibrated" submanifolds which the branes wrap \cite{K1, MS}. 
The RR flux, $F$, is dictated by supersymmetry to be 
$$F = d^\cja(e^{-2f}\, {\rm Im}\, \Phi_2) $$
and implicitly appears in the above equations, specifically in \eqref{eqc}, as the magnetic source equation $dF = \rho\,$.


\begin{table}[t]
\renewcommand{\arraystretch}{2.1}
\renewcommand{\tabcolsep}{.4cm}
\centering
\begin{tabular}{c|c|c}
Generalized Complex & IIB Complex  &  IIA Symplectic     \\
\hline
$\Phi_1$ & $\Om^{3,0}$ & $e^{i\,\om}$  \\
\hline
$\Phi_2$
& $e^{i\,\om}$ &$\Om^{3,0}$ \\
 \hline
$d^\cja$ & $d^c$ & $\dl$   \\
 \hline
$\dfrac{\ker d}{\im d d^\cja} $ & $\dfrac{\ker d}{\im d d^c}$ & $\dfrac{\ker d}{\im d \dl}$ \\
\hline
$\dfrac{\ker d d^\cja}{\im \pa_\cja + \im \bpa_\cja} $ & $\dfrac{\ker d\dc}{\im \pa +  \im \bpa}$ & $\dfrac{\ker d\dl}{\im \dpp + \im \dpm}$\\
\end{tabular}
\bigskip\bigskip

\caption{A comparison of the pure spinors, differential operators, and cohomologies between the generalized complex, complex, and symplectic cases.  The cohomologies defined on the space of  $\cj_1$-eigen-forms  $\CU^k_{\cj_1}$ (generalized complex), $(p,q)$-forms $\A^{p,q}$  (complex), and primitive forms $\CB^k$ (symplectic).}\label{firstops}
\end{table}


We now consider the linearized deformation of the above system generalizing the analyses of the two previous subsections.  The general linearized equations were worked out by Tomasiello in \cite{Tomasiello}.  Here we will perform a linearized variation of the almost generalized complex structure $\Phi_2 \to \Phi_2 + \d \Phi_2$ subjected to the following conditions: (i) keeping the integrable almost generalized complex structure represented by $\Phi_1$ fixed, i.e. $\d \Phi_1=0$: (ii) keeping the source current fixed, i.e. $\d \rho = 0$; (iii) keeping fixed the conformal factor, $\d f =0\,$.  

To begin, with two compatible almost generalized complex structures, $\d \Phi_2$ can be decomposed into eigen-forms of $(\cja, \cjb)$.  The corresponding eigenvalues are imaginary and we will denote them by $(i\,k_1,i\,k_2)$ where in six real dimensions, $-3 \leq k_1, k_2 \leq 3$.  In particular, $\Phi_1$ is a $(3i,0)$ and $\Phi_2$ is a $(0,3i)$ eigen-form.  We shall label the $k$-th eigen-forms of $\cja$ and $\cjb$, respectively, by $\CU^k_\cja$ and $\CU^k_\cjb$.  That the two structures are compatible implies $\d \Phi_2$ must be a zero eigen-form under $\cja$ (see e.g. \cite{Tomasiello}).    On the other hand, $\d\Phi_1=\d f = 0$, implies from \eqref{eqa} that $\d \Phi_2 \w \la ({\bar \Phi_2})$ vanishes.  But since ${\bar \Phi_2}$ is a $(0, -3i)$ eigen-form and as an infinitesimal variation, $\d \Phi_2\in U^3_\cjb \oplus U^{1}_\cjb\,$,  we therefore find that $\d \Phi_2$ must be an $(0,i)$ eigen-form.     

With a positive metric defined by the two compatible generalized structures $(\cj_1,\cj_2)$, we shall use it to define the Hodge star operator and take the inner product to be
\be\label{inprod}
(U_1, U_2) = \int_{X^6} e^{-2f} \left(\,U_1 \w  *  {\bar U_2}\,\right)_{top}
\ee
where $U_1$ and $U_2$ are generally sums of differential forms, or more precisely, spinors in $CL(6,6)$.   Though the $(i\,k_1, i\,k_2)$ eigen-forms are not eigen-forms of the Hodge star operator, they are eigen-forms of $*\la\,$.  In particular, acting on the $(0,i)$ eigen-form, $\d\Phi_2$, $* \la (\d\Phi_2) = -i \d\Phi_2\,$.  Hence, we have
\be\label{gcdual}
\Im \d\Phi_2 =  \frac{1}{2i} \left(\d\Phi_2 - \d {\bar \Phi_2}\right) = * \la
\frac{1}{2}\left(\d\Phi_2 + \d {\bar \Phi_2}\right)= *\la \left(\Re \d\Phi_2\right)~.
\ee
Thus, the variation of the generalized complex system with our stated constraints reduces to 
\be\label{gccond}
d\,\Re \d \Phi_2 =0~, \qquad  dd^\cja e^{-2f} *\la \left(\Re \d \Phi_2\right) =0~,
\ee
or alternatively,
\be\label{gccondd}
dd^\cja e^{-2f} \Im \d\Phi_2 =0~, \qquad d\, *\la \left(\Im \d\Phi_2\right)=0~.
\ee
These conditions which are harmonic type lead us to introduce the following two generalized complex cohomologies
$$H_{\pa_\cj+\bpa_\cj}^{k}(X) = \frac{\{U^k\in \CU_\cj^{k}| d\, U^{k} =0\}}{dd^\cj\,\CU_\cj^{k}}~,$$
and
$$H^{k}_{dd^{\cj}}(X) = \frac{\{U^k\in \CU_\cj^{k}| dd^\cj\, U_\cj^{k} =0\}}{{\pa}_\cj\,\CU_\cj^{k-1} + {\bpa}_\cj\,\CU_\cj^{k+1}}~.$$
with $\cj$ a generalized complex structure.  In above, we have used the decomposition of $d= {\pa}_\cj + {\bpa}_\cj$ which follows from $d: \CU_\cj^{k} \to \CU_\cj^{k+1} + \CU_\cj^{k-1}\,$ when $\cj$ is integrable \cite{Gua1}. 
These two cohomologies are natural extension of their complex and symplectic counterparts described in the previous subsections when situated in the generalized complex framework.  Just like in \cite{Sch,TY1}, their harmonic eigen-forms can be described as the solutions of fourth-order self-adjoint differential operators, $D^4_{\pa_\cj+\bpa_\cj}$ and $D^4_{dd^{\cj}}$.  Showing that these fourth-order operators are elliptic then demonstrate that the cohomologies are finite-dimensional. \footnote{Following the work of \cite{TY1}, the extension of the complex and symplectic cohomologies to the generalized complex case has also been independently written down by G. Cavalcanti and M. Gualtieri \cite{CG2}.}  

Thus, with the imposed conditions $\delta\Phi_1= \delta \rho = \delta f =0$ and with \eqref{gccond}-\eqref{gccondd}, we find that the linearized deformation of $\Re \delta \Phi_2$, or alternatively $\Im \delta \Phi_2$, can be parametrized by harmonic eigen-forms with a definite action on $\la$ (e.g. $\la (\Re \delta \Phi_2) = + \Re \delta \Phi_2$ or $\la (\Re \delta \Phi_2) = - \Re \delta \Phi_2\,$) \footnote{The choice of the sign under the action of $\la$ is affected by the type of orientifold sources that are present, see e.g. \cite[Appendix D]{Martucci}.} such that
$$\Re \delta\Phi_2 \in \CH^{0}_{\pa_\cja + \bpa_\cja}(X)\cap \Re \CU_\cjb^1~,$$
or 
$$\Im \delta\Phi_2 \in \CH^{0}_{dd^\cja}(X)\cap \Im \CU_\cjb^1~.$$
Let us add that in general, the requirement of $\d \rho =0$ can give obstructions to the above $\d \Phi_2$ deformations.

\section{Concluding Remarks}

The purpose of this paper has been twofold: (1) to show that the complex cohomologies of Bott-Chern and Aeppli, and the symplectic cohomologies of Tseng and Yau have application to counting massless modes in type II flux compactifications; (2) to extend the complex and symplectic cohomologies result within the generalized complex framework of type II theory which naturally lead us to two new generalized complex cohomologies.

It should be clear that the cohomologies we have emphasized here differ from the more standard cohomologies (e.g. de Rham cohomology) when the $d\dJ$-lemma (the generalized complex generalization of the $d\dc$-lemma of complex geometry) fails to hold.  This lemma is of course not a requirement of supersymmetry.  In fact, many simple type II ${N=1}$ supersymmetric flux backgrounds \cite{GMPT2} are built from nil-manifolds, which are generalized complex and also generically do not satisfy the $d\dJ$-lemma \cite{CG, Caval}.   For such backgrounds, it is possible to explicitly calculate the different cohomologies and see the differences in their dimensions.  For instance, in the simple complex nil-manifold background solutions, it is straightforward to compute and find that the dimension of the Dolbeault cohomology actually undercounts in certain examples the number of moduli fields when compared with the Bott-Chern and Aeppli cohomology.

As mentioned, the motivation of this work comes from the Maxwell equations.  Besides its relation to the space of Maxwell solutions, the de Rham cohomology also plays a role as the relative cohomology of the source current, $\rho_e$.  Indeed, one can ask given the ${N=1}$ supersymmetric equations, what cohomology describe the currents of the supersymmetric branes.     The equations naturally suggest the relative versions of the cohomologies we have highlighted here.  

The presence of branes sources present another subtlety which we have ignored.  Because branes are represented by singular currents in the equations, all geometrical quantities necessarily becomes singular on the support of the branes.  The type of cohomologies characterizing the moduli should rigorously be those with compact support and vanishing along the branes.  Such an approach has been discussed in \cite{Hu}.  

As mentioned, the equations above are general and hold for any supersymmetric configurations of branes and RR fluxes.  We have however ignored the NSNS fluxes, or the $H_3$ field.  In type II string theory without NSNS branes, $dH=0$, and the modification to the above generalized complex equations is simply replacing $d$ with $d_H=d- H\w\,$ \cite{GMPT1}.  The generalized complex cohomologies we have introduced can thus incorporate a non-zero $H$-flux by using $d_H$ operators instead of the exterior derivative $d$.

Finally, the analysis in this paper is at the level of linearized infinitesimal variation.  We have not delved into important issues such as obstructions to integrability, orientifold projections and open-string massless modes related to the branes.  Some of these issues have been explored in \cite{KM, Tomasiello} and especially \cite{Martucci} which made use of additionally physical consistency arguments related to four dimensional low-energy effective field theory.  Providing a full accounting of all the massless moduli from geometry will necessitate a deeper understanding of non-K\"ahler geometry than what is currently available.   In this paper, we have given yet another example that the mathematical tools involved in non-K\"ahler flux compactifications, in particular here cohomologies, are generally not identical to those in K\"ahler geometry and Calabi-Yau compactifications.  As geometries that are non-K\"ahler are much more diverse and flexible than that of K\"ahler Calabi-Yau, one expect that more refined tools will be required to characterize them.   Developing them will certainly help us gain deeper insights into vast regions of the still mysterious landscape of supersymmetric flux vacua.

\

\noindent{\bf Acknowledgements.~} 
We would like to thank G. Cavalcanti, F. Denef, M. Gualtieri, S. Hu, R. Minasian, A. Strominger, X. Sun, D. Waldram, and M. Zabzine for their interest and helpful comments.  We are especially indebted to A. Tomasiello for a number of  discussions on supergravity solutions and moduli fields.    The first author would also like to thank the kind support of the Department of Mathematics and the Center for the Fundamental Laws of Nature at Harvard University where a large portion of the work took place.  This work is further supported in part by NSF grants 0804454, 0854971, and 0937443.

\

\end{document}